\documentclass{singlecol-new}

\usepackage{url,natbib,stfloats}
\usepackage{mathrsfs}
\usepackage{graphicx}
\usepackage{soul}

\theoremstyle{TH}{

}

\theoremstyle{THrm}{

}

\theoremstyle{THhit}{

}

\makeatletter
\def\theequation{\arabic{equation}}

\makeatother

\usepackage{tcolorbox}
\tcbset{
  colback=gray!10,
  colframe=gray!60,
  boxrule=0.6pt,
  arc=2pt,
  left=6pt,right=6pt,top=4pt,bottom=4pt
}

\newcommand{\PreprintNoteIJDMMM}{%
\begin{tcolorbox}
\footnotesize
\textbf{Preprint (Author's Original Manuscript, AOM).}
The published version is available at: \texttt{https://doi.org/10.1504/IJDMMM.2022.125265}.
\end{tcolorbox}
}

\begin{document}%

\setcounter{page}{1}

\LRH{Author et~al.}

\RRH{Spanish cyberbullying detection model throughout deep learning}

\VOL{x}

\ISSUE{x}

\PUBYEAR{201X}

\BottomCatch

\CLline

\subtitle{}

\title{Detecting cyberbullying in Spanish texts throughout deep learning techniques}

\authorA{Pa\'ul Cumba-Armijos}
\affA{Digital School \\SEK International University \\Quito, Pichincha, Ecuador, Fax: +593-3994800 \\E-mail: pcumba.mti@uisek.edu.ec}

%
%
\authorB{Diego Riofr\'io-Luzcando}
\affB{Digital School \\SEK International University \\Quito, Pichincha, Ecuador Fax: +593-3994800 \\E-mail: diego.riofrio@uisek.edu.ec}
\authorC{Ver\'onica Rodr\'iguez-Arboleda}
\affC{Digital School \\SEK International University \\Quito, Pichincha, Ecuador Fax: +593-3994800 \\E-mail: veronica.rodriguez@uisek.edu.ec}

\authorD{Joe Carri\'on-Jumbo}
\affD{Digital School \\SEK International University \\Quito, Pichincha, Ecuador, Fax: +593-3994800 \\E-mail: joe.carrion@uisek.edu.ec}

%
%
%
%
%
%
%
\vspace{-7.0em}
\PreprintNoteIJDMMM
\vspace{5.5em}

\begin{abstract}
Recent recollected data suggests that it is possible to automatically detect events that may negatively affect the most vulnerable parts of our society, by using any communication technology like social networks or messaging applications. This research consolidates and prepares a corpus with Spanish bullying expressions taken from Twitter in order to use them as an input to train a convolutional neuronal network through deep learning techniques. As a result of this training, a predictive model was created, which can identify Spanish cyberbullying expressions such as insults, racism, homophobic attacks, and so on.

\end{abstract}


\KEYWORD{cyberbullying; deep learning; convolutional neuronal network; spanish; social networks}


\REF{to this paper should be made as follows: }

\begin{bio}
{Pa\'ul Cumba-Armijos obtained his Master's degree from the SEK International University in 2018, and his degree of Engineer in Computer Systems Engineering from the Polytechnic School of Chimborazo in 2012. He has worked on machine learning techniques for word processing in the field academic and has experience in agile software development techniques in the business field and multi platform development.\vs{9}

\noindent Diego Riofr\'io-Luzcando is the Director of the UISEK Digital School, he has a PhD in Software and Systems (UPM 2017), a Master in Software and Systems (UPM 2012) and a Computer Science and Computing Engineering (National Polytechnic School of Ecuador 2006). His main research interests include virtual education and training environments, smart tutoring systems, data science, and digital transformation.\vs{8}

\noindent Ver\'onica Rodr\'iguez-Arboleda. is the Coordinator of Basic and Generic Subjects of the UISEK. Teacher of the Master of Cybersecurity at Digital School. Accredited researcher at the Senescyt, Master in Business Administration. Master in University Teaching. Computer Science Systems Engineer. Research interest, Information Technologies applied to education, virtual environments, Cybersecurity.\vs{8}

\noindent Joe Carri\'on-Jumbo  received his PhD grade of Computer Science at Universitat Autónoma de Barcelona (UAB) in 2017. He received his Master degree from the Computer Architecture and Operating Systems (CAOS) at UAB in 2013. He worked for a private company for 10 years with four years as manager of the IT department. He is focused in studies related in areas related to database, data mining  and business intelligence.\vs{8}\\
}
\end{bio}

\maketitle

 \section{Introduction}

By allowing users to share and publish information in the different social networks freely, the problems especially to vulnerable groups such as children and teenagers have increased \citep{David-Ferdon2007}.

This phenomenon is aggravated when minors commit suicide, which has increased in recent years due to the appearance of cyberbullying, mainly because this type of violence does not end when the minor returns to his/her home, unlike the situations of face-to-face violence \citep{Girard2017}.

According to the Cybersecurity Research Center, about 28\% of students from middle and high schools around the United States have been victims of cyberbullying \citep{Patchin2016}. In Spain, The ANAR Fundation and Mutua Madrile\~na \citep{ANAR2016} affirm that one in four harassments that occurred in Spanish schools are produced using technological means. It was determined that 30.7\% of adolescents suffered from bullying through the internet in Latin America \citep{ESET2014}. For example, in some cities of Ecuador (Quito, Manta and Guayaquil), a high level of grooming and cyberbullying through social networks has been detected, about 27\% of teenagers have suffered marginalization, 46\% harassment, 17\% aggression and 10\% extortion \citep{Ortega2013}.

To prevent this type of aggression, it is essential to provide preventive control measures, for example identifying attacks on time \citep{Nandhini2015}. Data analytics techniques can be used as a prevention tool to detect this type of incidents.

Therefore, this work is the creation of a model using deep learning techniques that identify cyber bullying attacks in Spanish written texts. 83,400 tweets have been used to identify these attacks and compared to texts with no violence at all.

The structure of the remainder of the paper is as follows. Section \ref{sec:related_works} states relevant works in the fields of sentiment analysis in social networks and automatic cyber bullying prediction. In the Section \ref{sec:model} there will be a description of the proposed model which would detect cyber bullying in texts. Section \ref{sec:model_val} reports model validation which details the method observed in this study and discusses its results. The last section \ref{sec:conclusion} outlines the conclusions of this research and some takes a look at further work within its field.

\section{Related Work}\label{sec:related_works}
The related work is divided into three sections:

Section \ref{ssec:Sentiment} briefly presents the main goals, behavior analysis in social networks, especially how to interpret the sentiments in texts through the use of natural language processing (Spanish). Section \ref{ssec:CyberbullingSN} summarizes the reviewed literature about social networks and Cyberbullying published on specialized journals. Finally, Section \ref{ssec:Prediction} focuses on the use of machine and deep learning to create prediction models that can detect cyberbullying.

\subsection{Sentiment analysis in social networks}\label{ssec:Sentiment}
Historically, the study of human behavior has been approached with great interest from several disciplines, and this kind of researches is currently being carried out within social networks. Some authors highlight the importance of these services since they allow to extract patterns of a large amount of data to understand the behavior of the users in order to satisfy their needs in the consumption of information \citep{Hu2012}. This work extends this idea by focusing in a real context based on Social networks with specific needs related to a social problem.

 For example, some researches are used to understand human sociological behavior through a probabilistic and statistical model \citep{Zhang2011}, in order to apply this hypothesis text messages have been gathered and classified manually to ensure the quality of the model. Other works use a natural language analysis in order to identify the personality, social situation and interpersonal relationships that are established among individuals in a social network, \citep{Pennebaker2002}. Instead of identifying personal as well as social features, this work purposes classifying expressions for applying this model. There are researches that classify the main aspects that affect human aging based on data mining techniques \citep{Squicciarini2017}. Thus, this work applies techniques to a specific part of psychological human aspects which is the way the humans express their sentiments by analyzing the natural language.

Data analysis in social networks can also be used as a mean to improve business, for example to identify rumors in business Twitter accounts \citep{Kaya2019}, in contrast this work uses the messages of the accounts and the accounts an tried anonymously, or to create customized advertising to the needs of the buyer \citep{Schiaffino2009}. Instead of classifying the user accounts to raise actions, this work classifies messages in an offline way and create models to classify future messages. In this area, sentiments behind comments of users in social networks can be analyzed to measure the loyalty and interests of the clients with respect to any brand or product \citep{Neri2012}, which can be obtained by categorizing opinions (positive or negative) \citep{Baracho2012}.

Machine learning is one of the techniques that can help analyzing data in social networks that can be implemented. For example, to classify the sentiments in positive, negative or neutral depending on the user’s opinion about other people, organizations, events, products, services, places, ideas, among others \citep{Beigi2016,Jayasanka2014}. In works where deep learning is the technique used, the purpose of the models created where the same as those mentioned above, with the difference that the type of neuronal network adopted was the Convolutional Neural Network \citep{Lu2017,Severyn2015}.

\subsection{Cyberbulling and social networks}\label{ssec:CyberbullingSN}

In order to identify relevant studies about cyberbullying, a search of indexed journals was carried out. Two search criteria were used, the first one about cyberbullying in specialized data mining journals. The second group of searches on specialized magazines in social networks.

Criterion 1 Journals on Data Mining:
Criterion 1 Journals on Data Mining: The journals indexed as Q1, Q2, Q3 and Q4 in the SJR ranking were searched. The title and area of the journal should include the phrases “data mining”. In total, 12 journals were found, in the Q1 group 8 journals were identified, one in Q2, one in Q3 and two journals in Q4. In the 12 magazines with a direct focus on Data mining, articles on cyberbullying were searched. The result was 4 articles.

Criterion 2 Magazines on Social Networks:
Magazines Q1 to Q4 that include “Social Network” in their title were searched. 34 journals were identified among all quartiles. In these journals, articles that include the word cyberbulling in the Title, Abstract and keywords were searched. The result was 67 articles.

In the total of 71 articles, those ones that perform analysis or detection in the content related to cyberbullying were searched. Through a review of the abstract of the articles, 5 articles that propose models of content analysis for cyberbullying detection were identified.

It can be seen in \citet{silva2018bullyblocker} that they develop an automated model to identify and measure the degree of cyberbullying on Facebook. This research uses the Twitter Social network so it allows us to analyze a different group of users instead of Facebook. In the article \citet{raisi2018weakly} propose a weakly supervised model by means of vocabulary analysis using machine learning techniques which have been evaluated on Twitter, Ask.fm and Instagram networks. This work uses Neural Networks instead and it tries to improve the accuracy for a specific social area which is the cyberbullying behavior. 

The analysis of social networks based on the identification of positive and negative behavior is studied by \citet{Squicciarini2017}, in the article the complexity of the models is analyzed due to the high volume of traffic in the networks and their high exchange rate, positive and negative sentiments can be extended to several areas or topics and uses generic terminology. On the other hand, this work aim is to focus on Cyberbullying attacks as a main goal. 

Some networks that are not currently in use have been studied by \citet{rafiq2016analysis}, in this case the authors analyze the video content of the extinct Vine network by downloading and tagging videos to model and content. The tagging technique is very useful to validate a model, the same technique in order to create a Corpus will be applied. The video tagging strategy to search similar images, audios and videos is also proposed by \citet{soni2018see}, who use a corpus created by \citet{rafiq2016analysis} to analyze behavior in text and audios of the videos. The 66-remaining works analyze the presence of cyberbullying in social networks as per in general content.

\subsection{Cyberbullying Prediction}\label{ssec:Prediction}
As mentioned above, machine learning is used to detect sentiments in texts based on a given model. In this way some authors try to identify who?, what?, why?, where? and when? An episode of bullying occurs \citep{Bellmore2015}.  Proposed model can be used to the same goal and compare the result with other models. In order to accomplish this goal, 9,764,583 bullying posts were compiled from Twitter to obtain a classification on different categories: Yes or No (is cyberbullying), the role of the author (who), the way bullying is carried out (what), the post typology (why).

Specifically, for the analysis of feelings and the detection of cyberbullying in social networks in Spanish language, machine learning was used in the development of a Bayesian classifier, which was trained with a set of data obtained from the Faculty of Exact Sciences of the University of San Agustín in Buenos Aires \citep{Mercado2018}.

Thus, two researches for the detection of cyberbullying can be highlighted. The objective of the first was to identify global polarities as positive, negative and neutral from texts in Japanese \citep{Ptaszynski2017},, for which the authors had to use morphological analyzers to transcribe the texts into English, it is very important to apply this kind of models to other languages like Spanish. And the second research compares results of machine learning techniques with those of deep learning to predict cyberbullying in English, obtaining better results for Convolutional Neural Network models \citep{Agrawal2018}.

Most researches that focus on cyberbullying prediction have been carried out in languages other than Spanish (Arabic, English), being Convolutional Neural Network, the most common method used \citep{Bayari2021}. This predilection for deep learning techniques may occur due to the fact that it has best-in-class performance on problems in text classification domain \citep{Altinel2018}.

\section{Cyberbullying detection model}\label{sec:model}
\subsection{Model architecture}
The proposed architecture is structured in phases (Figure \ref{fig:architecture}), which allows to establish a controlled and independent flow of information. Each phase determines one or several processes used for the preparation of the neural network:
\begin{figure}
    \centering
    \includegraphics[width=\columnwidth]{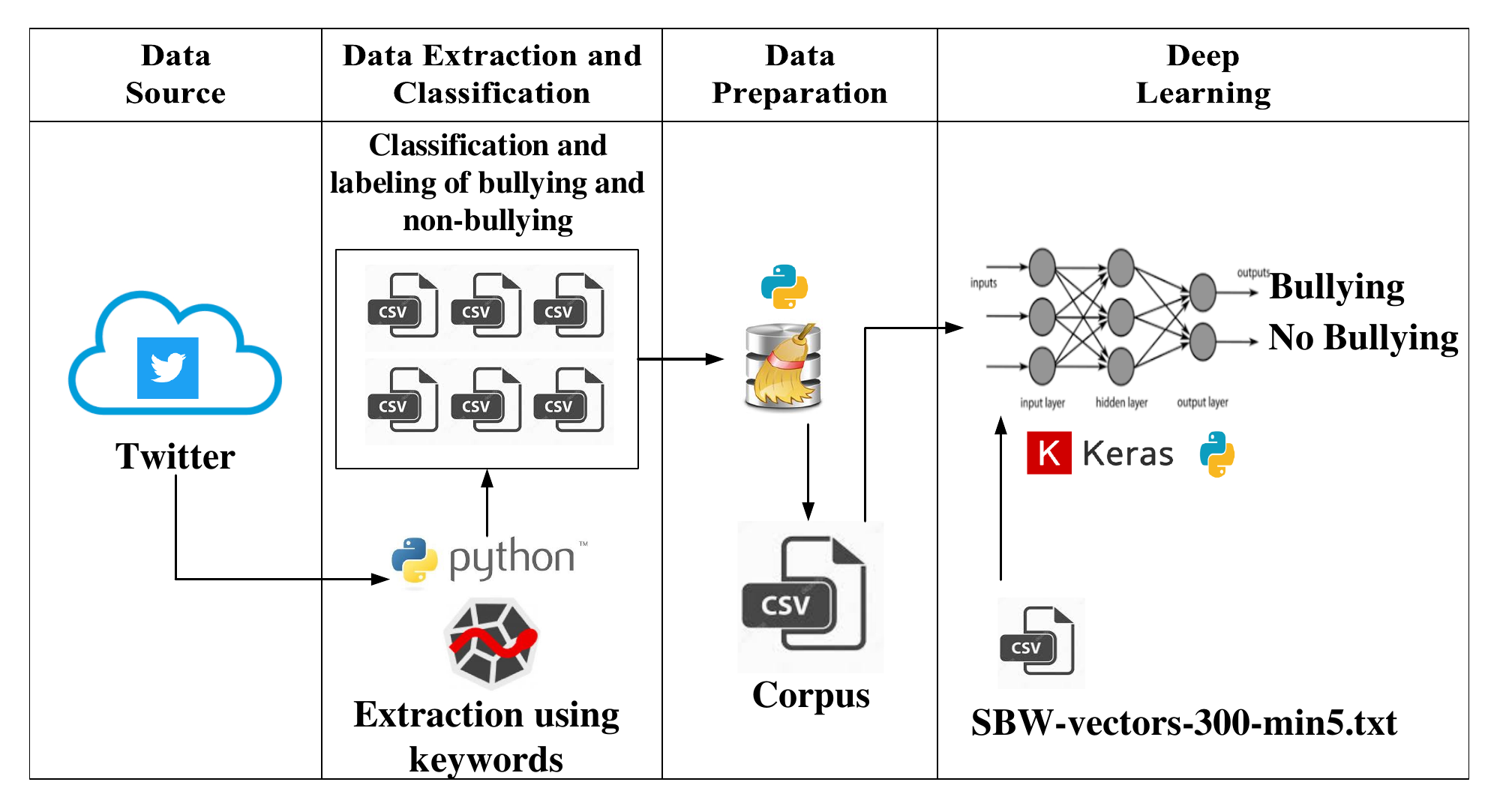}
    \caption{Cyberbullying detection mode architecture}
    \label{fig:architecture}
\end{figure}


As it can be appreciated in \ref{fig:architecture}, the first phase is responsible for collecting the information to train the model, for which 83,400 tweets were obtained. The second phase defines the scheme of data extraction, this process was developed in a Python script which interacts with the social network Twitter through its API. 

In the data preparation phase, each tweet was manually classified and labeled with zero (no bullying) or one (bullying), depending on the context and keywords \citep{DelRey2002}, resulting on 16,247 bullying labeled tweets and 67,153 no bullying labeled tweets. In this phase, data quality has been improved in order to avoid words or symbols that may cause noise in model training. For this, a Python script was also developed in which the following actions were executed:
\begin{itemize}
\item Convert text to lowercase.
\item Delete all special characters used in each tweet. Only letters remain.
\item Remove references of external links ("http", "https", "www").
\item Eliminate mentions to other users "@user".
\item Remove stop words.
\end{itemize}

Finally, a Convolutional Neural Network was trained using these preprocessed tweets. For this, a Python script was developed by means of Keras 3 module. The Convolutional Neural Networks are an excellent classifier of sentences, because this type of networks can capitalize the distribution of words in a sentence by converting them into a vector matrix \citep{Kim2014}.

\subsection{Data extraction}
For the search and extraction of the tweets the labels that young students between 8 and 14 years of age used to denominate the phenomenon of bullying were considered \citep{DelRey2002}, to search for texts of verbal aggression (bullying), to search for texts of verbal aggression (bullying). Labels such as insults, racist attacks and serious verbal violence (Table \ref{tab:bullyingwords}). At the same time, opposite words to the violent ones were used to find harmless tweets (Table \ref{tab:nobullyingwords}).


\begin{table}[h]
\renewcommand{\arraystretch}{1.3}
\caption{Example of keywords for bullying expressions (Spanish)}
\label{tab:bullyingwords}
\centering
\begin{tabular}{|c c c|}
\hline
care asno & care culo & care huevo \\
eres regalada & homosexual & gay \\
maricon & mamerto & mojigato \\
nerd & ojala mueras & orejas elefante \\
te den culo & te odio & tetas chuecas \\
care simio & care verga & eres gorda \\
borrego & maldito & marica \\
narizon & negro & apestoso \\
parece cerda & parece marrana & parece marrano \\
eres deprimente & eres discapacitado & eres patetico \\
\hline
\end{tabular}
\end{table}

\begin{table}[ht]
\renewcommand{\arraystretch}{1.3}
\caption{Example of keywords for no bullying expressions (Spanish)}
\label{tab:nobullyingwords}
\centering
\begin{tabular}{|c c c|}
\hline
gusta & me gustas & linda \\
gustoso & eres mejor & lindo \\
pago entradas & animo & enamorarse \\
hola & amiga & frio \\
amigo & cuidate & calor \\
te quiero & gustoso conocerte & tener autoestima \\
te amo & como estas & vive feliz \\
\hline
\end{tabular}
\end{table}

\subsection{Creating the model}
\begin{figure}[ht]
 \centering
 \includegraphics[width=\columnwidth]{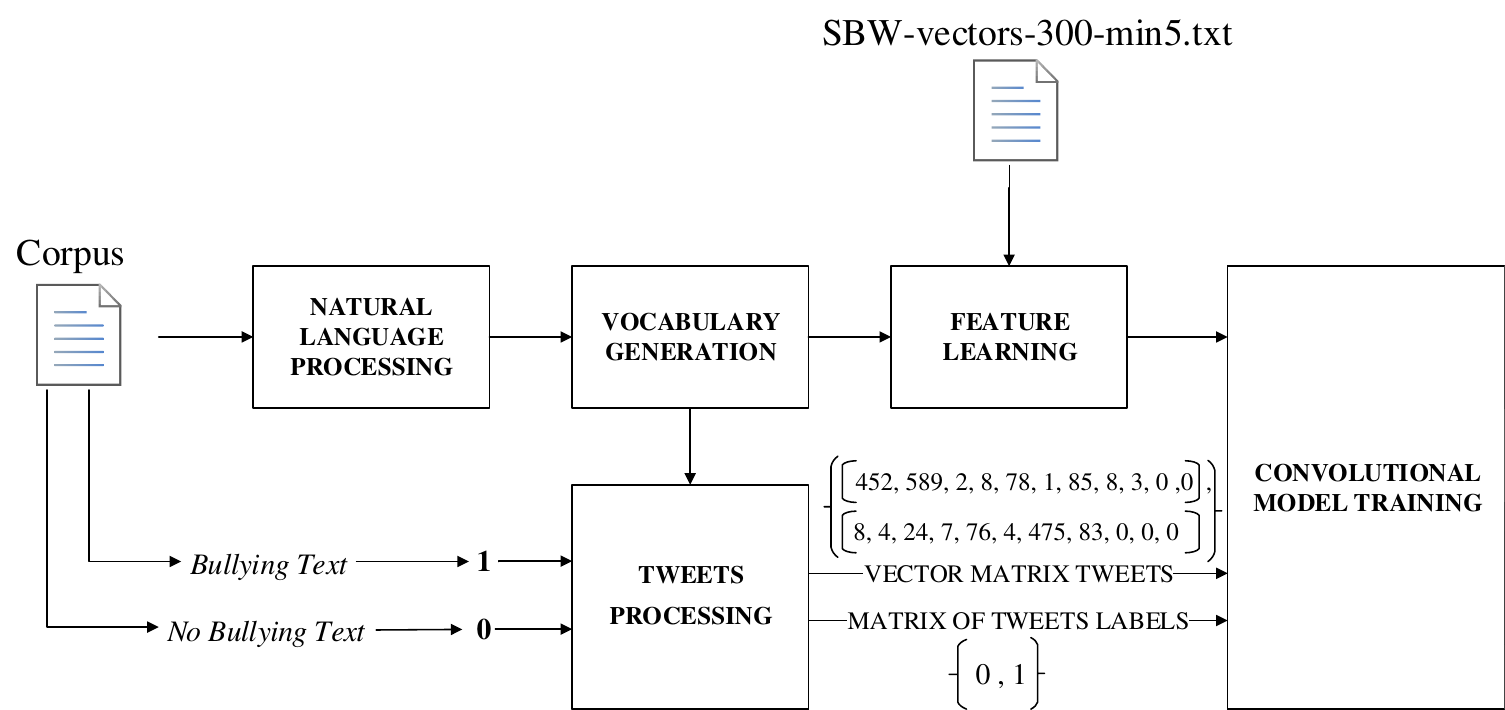}
 \caption{Generation process of the model}
 \label{fig:modelprocess}
\end{figure}

A flow with five tasks was defined in order to create the model (Figure \ref{fig:modelprocess}), based on the traditional machine learning process \citep{Raschka2015}.

\subsubsection*{Natural language processing}
The frequency distribution of the words corpus is obtained through the NLTK library.

\subsubsection*{Vocabulary generation}
This vocabulary is created based on the frequency distribution, by implementing the most common words and then an indexed matrix. This matrix starts with the most frequent word assigning the number one (1) as an index, the second most frequent word is assigned the number two (2) as an index, and so on.

\subsubsection*{Tweets processing}
This process decomposes the corpus of the training data set into three vectors.
\begin{itemize}
\item Tweets. Stores the values of the indexes, which represent the words of the corpus based on the frequency. 
\item Labels. Stores the values of the labels: one (1) if is bullying and zero (0) if it is not.
\item Ids. Stores the ids of each tweet.
\end{itemize}

\subsubsection*{Feature learning}
For this, an incrustation of words in Spanish was used, words vectorially represented through a linguistic model consolidated in a file (SBW-vectors-300-min5.txt) from different resources published on the web, and made up of 1.4 billion words in Spanish \citep{Cardellino2016}. This model establishes analogies between words and their proximity relationships through a vectoral representation, and it was created by using the word2vec algorithm \citep{Mikolov2013}.

\subsubsection*{Model training}
As stated above, a Convolutional Neural Network was used to train the model. This network receives the corpus of preprocessed tweets as input and emits predictions whether a text is bullying or not as output. 

\section{Model validation}\label{sec:model_val}
\subsection{Method}\label{sec:method}
The validation of the model was carried out in two parts. First, a validation of the consistency of the corpus and then a cross-validation of the model to find the accuracy of its predictions.

\subsubsection*{Corpus validation}
As previously stated, the corpus was structured from a total of 83,400 tweets that describe texts that have bullying and non-bullying signs. Therefore, it is important to establish that the words used to create the corpus are consistent and satisfy the minimum requirements to represent expressions of interpersonal communication.

For this, an analysis of the linguistic distribution of the words in tweets used for the Corpus was carried out, as stated by Zipf \citep{Zipf1949}. This law expresses that people use a reduced number of words most of the time, while the vast majority of words are used very rarely in the linguistic communication. This use of words follows a distribution that can be represented by:
\[f(r)=\alpha \frac{1}{r^\alpha}; \alpha \approx 1\]

Where \(f(r)\) represents the frequency that it is most used and \(\alpha\) is a number that approaches one. This equation means that the frequency of each word is inversely proportional to its location range according to its frequency, so, the most frequent word will use twice more than the second most frequent word and three times more than the most used words, and so on.

\subsubsection*{Model cross-validation}
For this validation, the total data set was divided into two subsets (training and test sets), after that the validation was carried out in four random iterations. The proportion to perform the data split was 90\% for the training set (75059 tweets) and 10\% for the test set (8341tweets).

Once the different data sets were obtained, the convolutional network was trained during each iteration with eight epochs.

In each iteration at the end of each epoch a checkpoint was established, storing a model as a result of each epoch. That is to say, eight epochs were executed for each training iteration and a model was stored at each checkpoint, which generated a total of 32 models with their respective accuracy and loss.

These metrics were used to select the best generated model in each training iteration and for each epoch. Thus, in terms of the model that has the highest precision and the least loss.

Once the best model of each iteration was selected, a validation of how well that model predicts the data from the test set was made, with which results of success and failure were obtained for each prediction.

\begin{figure}[b]
 \centering
 \includegraphics[width=0.69\columnwidth]{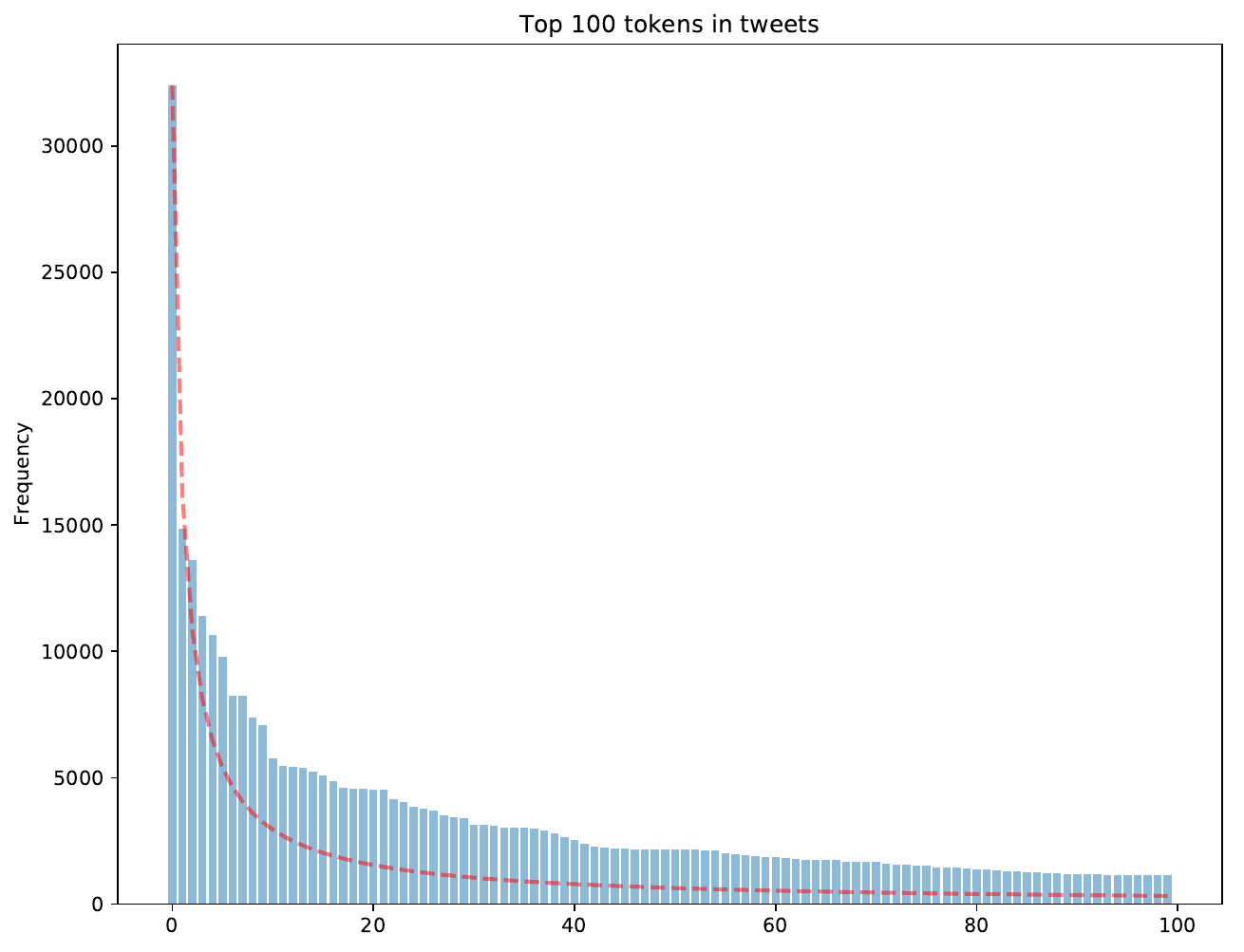}
 \caption{Top 100 of the frequency distribution of the words in the corpus}
 \label{fig:zipdistribution}
\end{figure}

\subsection{Results}
\subsubsection*{Corpus validation}
Figure \ref{fig:zipdistribution} shows the distribution according to the frequency range of the words of the bullying and non-bullying tweets used in the corpus, while red dotted line represents the Zipfian distribution.

\begin{figure}[h]
 \centering
 \includegraphics[width=0.7\columnwidth]{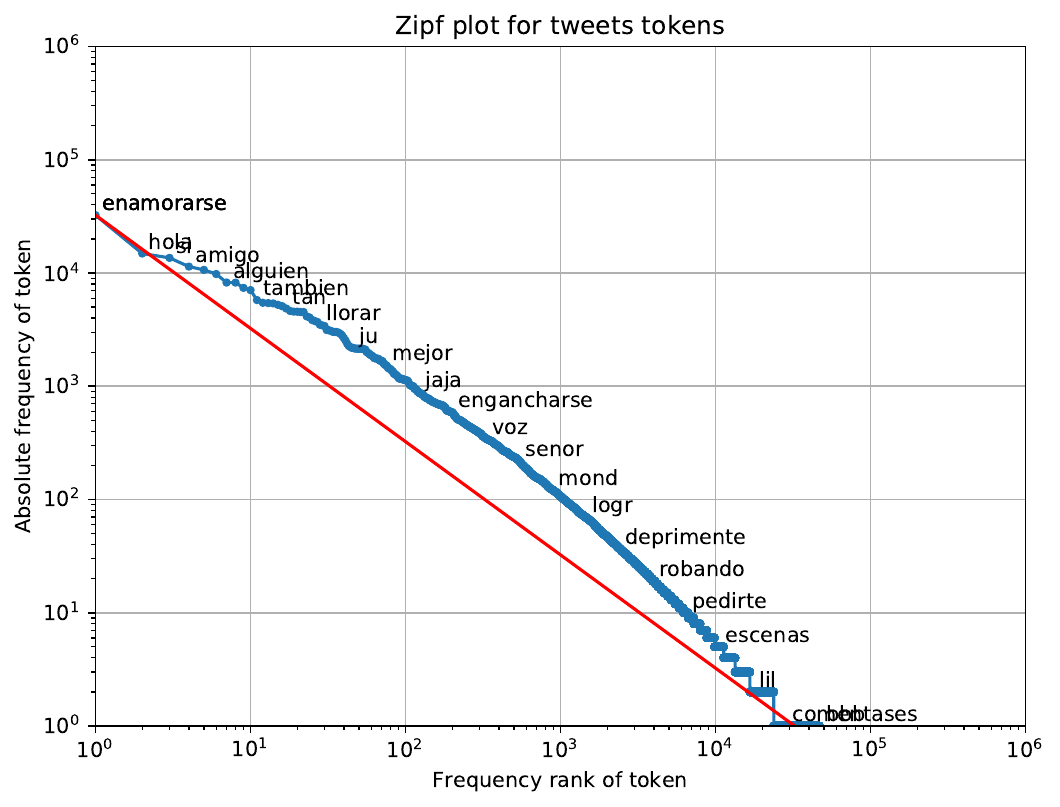}
 \caption{Words distribution vs a Zipf distribution}
 \label{fig:zipzdistaprox}
\end{figure}

Another representation to verify that the distribution of the corpus complies with Zipf’s law is shown \ref{fig:zipzdistaprox}, which represents the frequency range of the words used in the corpus vs. its absolute frequency.

When plotting the frequencies of the words in the plane, an almost linear curve approximates to the red line can be observed, which represents the distribution of Zipf.

\subsubsection*{Model cross-validation}
As stated in the section \ref{sec:method}, eight epochs were used for each training iteration of the model and for each iteration the best model was selected. The accuracy and loss of each of these selected models are presented in Table \ref{tab:modelsepoch}.

\begin{table}[h]
\renewcommand{\arraystretch}{1.3}
\caption{Model selected for each training iteration}
\label{tab:modelsepoch}
\centering
\begin{tabular}{c c c c}
\hline
Training Iteration & Selected Model & Accuracy & Lost\\
\hline
1	&	Epoch 8	&	99.80\%	&	0.006\\
2	&	Epoch 8	&	99.80\%	&	0.005\\
3	&	Epoch 8	&	99.80\%	&	0.006\\
4	&	Epoch 8	&	99.80\%	&	0.005\\
\hline
\end{tabular}
\end{table}

With the best selected model, the cross-validation was carried out to prove how well the model predicts a text with and without bullying. Results that are presented in Table \ref{tab:crossvalidation}, where the model hits a 98.85\% of the time average prediction.

\begin{table}[h]
\renewcommand{\arraystretch}{1.3}
\caption{Results of success and failure in the prediction of test set for each iteration.}
\label{tab:crossvalidation}
\centering
\begin{tabular}{c c c}
\hline
Iteration & Success Percentage & Fail Percentage\\
\hline
1	&	98.96\%	&	1.04\%	\\
2	&	98.84\%	&	1.16\%	\\
3	&	98.92\%	&	1.08\%	\\
4	&	98.67\%	&	1.33\%	\\
\multicolumn{1}{r}{\textbf{Average}} & \textbf{98.85\%} & \textbf{1.15\%}\\
\hline
\end{tabular}
\end{table}

\subsection{Discussion}
Even though in Figure \ref{fig:zipdistribution} it can be seen that the use of the words in the corpus resembles that distribution, it can also be observed in Figure \ref{fig:zipzdistaprox} that there is a deviation of the corpus distribution curve from the expected distribution, which means that in the corpus some words occur with a higher frequency range. 
This may mean that the tweets used for the corpus have more frequent words than normal expressions. This may be due to the fact that keywords were used to obtain the corpus and these are the most frequent words in the corpus.

In the cross-validation, the results obtained allow us to show that the models in the different validation iterations do not have a high variability among them. Besides that, in each one of the iterations and on average a high percentage of precision is presented.

\section{Conclusions and Future Work}\label{sec:conclusion}

This research presents a contribution to the development of systems that prevent and alert cases of cyberbullying, this is accomplished by taking advantage of the model generated from a corpus with Spanish expressions labeled as bullying and non-bullying.

In the extraction of data from Twitter, the importance of retrieving focused data using keywords is highlighted, which allow us to obtain the expressions that represents concrete situations of interpersonal violence. In addition, this process allowed us to obtain a large amount of information, which was an essential pillar for the preparation of the model.

The training process of the Convolutional Neuronal Network was carried out efficiently using deep learning techniques, since the convolution layers of the neural network facilitated the extraction and the understanding of the characteristics of the linguistic relationship of the words in the expressions that had signs of bullying, thanks to the help of a pre-trained vector matrix with Spanish words. In this process the data preparation that was made to the corpus was of essence, since this guaranteed that the processing of the model was efficient in time and quality of training.

The use of the cross-validation technique allowed us to verify that the models had in average a very high precision (98.19\%) when predicting the test data.

Inclusion of data from other sources like other social networks or blogs will be needed in order to generate a larger corpus for a future work with the objective of feeding the neural network with more linguistic characteristics that can be obtained from other expressions used in Spanish, in order to generate a model that can improve the prediction values.




\begin{thebibliography}{}

\bibitem[Agrawal and Awekar, 2018]{Agrawal2018}
Agrawal, S., Awekar, A.: ``Deep learning for detecting cyberbullying across
  multiple social media platforms''; European Conference on
  Information Retrieval, Springer, (2018), 141-153.

\bibitem[Alt{\i}nel and Ganiz, 2018]{Altinel2018}
Alt{\i}nel, B. and Ganiz, M. C.: ``Semantic text classification: A survey of past and recent advances''; Information Processing \& Management, 54(6), (2018), 1129-1153.
  
\bibitem[ANAR and Mutua Madrile\~na, 2016]{ANAR2016}
ANAR, F. and Mutua Madrile\~na, F.: ``Estudio sobre el ciberbullying
  seg\'un los afectados''; ANAR Fundation, Mutua Madrile\~na
  Fundation, Madrid, Tech. Rep., (2016). [Online]. Available:
  \url{www.fundacionmutua.es}
  
\bibitem[Baracho et~al., 2012]{Baracho2012}
Baracho, R.~M.~A., Silva, G.~C., Ferreira, L.~G.~F.: ``Sentiment Analysis in
  Social Networks: a Study on Vehicles.''; ONTOBRAS-MOST, (2012), 132-143.

\bibitem[Bayari and Bensefia, 2021]{Bayari2021}
Bayari, R. and Bensefia, A.: ``Text mining techniques for cyberbullying detection: state of the art''; Advances in Science, Technology and Engineering Systems, 6 (1), (2021), 783-790.

\bibitem[Beigi et~al., 2016]{Beigi2016}
Beigi, G., Hu, X., Maciejewski, R., Liu, H.: ``An overview of sentiment
  analysis in social media and its applications in disaster relief'';
  Sentiment analysis and ontology engineering, Springer, (2016), 313-340.

\bibitem[Bellmore et~al., 2015]{Bellmore2015}
Bellmore, A., Calvin, A.~J., Xu, J.~M., Zhu, X.: ``The five W's of
  “bullying” on Twitter: Who, what, why, where, and when'';
  Computers in human behavior, 44, (2015), 305-314.
  
 \bibitem[Cardellino, 2016]{Cardellino2016}
Cardellino, C.: ``Spanish Billion Words Corpus and
  Embeddings''; (2016). [Online]. Available:
  \url{https://crscardellino.github.io/SBWCE/}
  
\bibitem[Del Rey et~al., 2002]{DelRey2002}
{Del Rey}, R., Genevat, R., Ruiz, R.~O.: ``Etiquetas verbales en el
  vocabulario de docentes, padres y madres para nominar el fen{\'{o}}meno
  bullying''; Revista electr{\'{o}}nica interuniversitaria de
  formaci{\'{o}}n del profesorado, 5, 4, (2002), 11.
  
\bibitem[David-Ferdon and Hertz, 2007]{David-Ferdon2007}
David-Ferdon, C., Hertz, M.~F.: ``Electronic media, violence, and
  adolescents: An emerging public health problem''; Journal of
  Adolescent Health, 41, 6, (2007), S1-S5.

\bibitem[ESET, 2014]{ESET2014}
ESET: ``Reporte Sustentabilidad''; Tech. Rep., (2014). [Online]. Available:
  \url{http://www.eset-la.com/micrositios/responsabilidad-social/pdf/eset-reporte-resumido.pdf}

\bibitem[Galarsi et~al., 2011]{Galarsi2011}
Galarsi, M.~F., Medina, A., Ledezma, C., Zanin, L.: ``Comportamiento, historia
  y evoluci{\'{o}}n''; Fundamentos en humanidades, 12, 24, (2011), 89-123.

\bibitem[Girard, 2017]{Girard2017}
Girard, G.: ``El suicidio en la adolescencia y en la juventud''; Revista
  de Formaci{\'{o}}n Continuada de La Sociedad Espa{\~{n}}ola de Medicina de La
  Adolescencia, (2017).

\bibitem[Heidemann et~al., 2012]{Heidemann2012}
Heidemann, J., Klier, M., Probst, F.: ``Online social networks: A survey of a
  global phenomenon''; Computer networks, 56, 18, (2012), 3866-3878.

\bibitem[Hu and Liu, 2012]{Hu2012}
Hu, X., Liu, H.: ``Text analytics in social media''; Mining text
  data, Springer, (2012), 385-414.
  
\bibitem[Jayasanka et~al., 2014]{Jayasanka2014}
Jayasanka, R., Madhushani, M.~D.~T., Marcus, E.~R., Aberathne, I., Premaratne, S.~C.: 
``Sentiment analysis for social media''; International
  Journal of Advanced Research in Computer Science and Software Engineering, (2014).

\bibitem[Kaya and Alhajj, 2019]{Kaya2019}
Kaya, M., Alhajj, R.: ``Influence and Behavior Analysis in Social
  Networks and Social Media''; Springer, (2019).

\bibitem[Kim, 2014]{Kim2014}
Kim, Y.: ``Convolutional neural networks for sentence classification''; 
  arXiv preprint arXiv:1408.5882, (2014).

\bibitem[Lu et~al., 2017]{Lu2017}
Lu, Y., Sakamoto, K., Shibuki, H., Mori, T.: ``Are deep learning methods
  better for twitter sentiment analysis''; Proc.  23rd
  annual meeting of natural language processing (Japan), (2017), 787-790.

\bibitem[Mercado et~al., 2018]{Mercado2018}
Mercado, R.~N.~M., Chuctaya, H.~F.~C., Gutierrez, E.~G.~C.: ``Automatic
  Cyberbullying Detection in Spanish-language Social Networks using Sentiment
  Analysis Techniques''; International Journal of Advanced Computer
  Science and Applications, 9, 7, (2018).

\bibitem[Mikolov et~al., 2013]{Mikolov2013}
Mikolov, T., Chen K., Corrado, G., Dean, J.: ``Efficient estimation of word
  representations in vector space''; arXiv preprint arXiv:1301.3781,
  (2013).

\bibitem[Nandhini and Sheeba, 2015]{Nandhini2015}
Nandhini, B.~S. and Sheeba, J.: ``Online Social Network Bullying Detection Using
  Intelligence Techniques''; Procedia Computer Science, 45, (2015), 485-492. [Online]. Available:
  \url{https://www.sciencedirect.com/science/article/pii/S187705091500321X}

\bibitem[Neri et~al., 2012]{Neri2012}
Neri, F., Aliprandi, C., Capeci, F., Cuadros, M., By, T.: ``Sentiment analysis
  on social media''; 2012 IEEE/ACM International Conference on
  Advances in Social Networks Analysis and Mining, IEEE, (2012), 919-926.

\bibitem[{Ortega Mora}, 2013]{Ortega2013}
{Ortega Mora}, A.~C.: ``Manifestaciones de la agresi{\'{o}}n verbal entre
  adolescentes escolarizados''; (2013).
  
\bibitem[Patchin, 2016]{Patchin2016}
Patchin, J.: ``Summary of Our Cyberbullying Research (2004-2016)''; (2016).
  [Online]. Available:
  \url{https://cyberbullying.org/summary-of-our-cyberbullying-research}

\bibitem[Pennebaker, 2002]{Pennebaker2002}
Pennebaker, J.~W.: ``What our words can say about us: Toward a broader language
  psychology''; Psychological Science Agenda, 15, 1, (2002), 8-9.

\bibitem[Ptaszynski et~al., 2017]{Ptaszynski2017}
Ptaszynski, M., Eronen, J.~K.~K., Masui, F.: ``Learning deep on cyberbullying
  is always better than brute force''; IJCAI 2017 3rd Workshop on
  Linguistic and Cognitive Approaches to Dialogue Agents (LaCATODA 2017),
  Melbourne, Australia, (2017), 19-25.

\bibitem[Raschka, 2015]{Raschka2015}
Raschka, S.: ``Python machine learning''; Packt Publishing Ltd, (2015).

\bibitem[Schiaffino and Amandi, 2009]{Schiaffino2009}
Schiaffino, S., Amandi, A.: ``Intelligent user profiling''; Artificial Intelligence An International Perspective, Springer, (2009), 193-216.

\bibitem[Severyn and Moschitti, 2015]{Severyn2015}
Severyn, A., Moschitti, A.: ``Twitter sentiment analysis with deep
  convolutional neural networks''; Proc.  38th
  International ACM SIGIR Conference on Research and Development in Information
  Retrieval, ACM, (2015), 959-962.

\bibitem[Squicciarini et~al., 2017]{Squicciarini2017}
Squicciarini, A., Rajtmajer, S., Griffin, C.: ``Positive and negative
  behavioral analysis in social networks''; Wiley Interdisciplinary
  Reviews: Data Mining and Knowledge Discovery, 7, 3, (2017), e1203.
  
\bibitem[Zhang et~al., 2011]{Zhang2011}
Zhang, H., Dantu, R., Cangussu, J.~W.: ``Socioscope: Human relationship and
  behavior analysis in social networks''; IEEE Transactions on Systems,
  Man, and Cybernetics-Part A: Systems and Humans, 41, 6, (2011), 1122-1143.

\bibitem[Zipf, 1949]{Zipf1949}
Zipf, G.~K.: ``Human behavior and the principle of least effort''; (1949).

\bibitem[Raisi and Huang, 2018]{raisi2018weakly}
Raisi, E., Huang, B. (2018). Weakly supervised cyberbullying detection with participant-vocabulary consistency. Social Network Analysis and Mining, 8(1), 38.

\bibitem[Rafiq et al., 2016]{rafiq2016analysis}
Rafiq, R. I., Hosseinmardi, H., Mattson, S. A., Han, R., Lv, Q., Mishra, S. (2016). Analysis and detection of labeled cyberbullying instances in Vine, a video-based social network. Social network analysis and mining, 6(1), 88.

\bibitem[Silva et~al., 2018]{silva2018bullyblocker}
Silva, Y. N., Hall, D. L., Rich, C. (2018). BullyBlocker: toward an interdisciplinary approach to identify cyberbullying. Social Network Analysis and Mining, 8(1), 18.

\bibitem[Soni and Singh, 2018]{soni2018see}
Soni, D., Singh, V. K. (2018). See no evil, hear no evil: Audio-visual-textual cyberbullying detection. Proceedings of the ACM on Human-Computer Interaction, 2(CSCW), 1-26.

\end{thebibliography}


\end{document}